\documentclass[conference]{IEEEtran}
\IEEEoverridecommandlockouts
\usepackage{cite}
\usepackage{amsmath,amssymb,amsfonts}
\usepackage{graphicx}
\usepackage{textcomp}
\usepackage{xcolor}

\newcommand{\nRC}{\mbox{\textit{nRC}}}
\newcommand{\nCCDAB}{\mbox{\textit{nCCDAB}}}

\usepackage{svg}
\svgsetup{inkscapeversion=1}
\usepackage{stfloats}
\def\BibTeX{{\rm B\kern-.05em{\sc i\kern-.025em b}\kern-.08em
    T\kern-.1667em\lower.7ex\hbox{E}\kern-.125emX}}
\begin{document}

\title{RH+: Row-Hit-Optimized Scheduling for PIM-based LLM Inference}

\author{
\IEEEauthorblockN{Yongchan Jung\IEEEauthorrefmark{1},
Shafayat Mowla Anik\IEEEauthorrefmark{2},
Byeong Kil Lee\IEEEauthorrefmark{2},
and Jeeho Ryoo\IEEEauthorrefmark{1}}
\IEEEauthorblockA{\IEEEauthorrefmark{1}\textit{Fairleigh Dickinson University},
Vancouver, BC, Canada}
\IEEEauthorblockA{\IEEEauthorrefmark{2}\textit{University of Colorado Colorado Springs},
Colorado Springs, CO, USA}
\IEEEauthorblockA{Emails: y.jung@student.fdu.edu, sanik@uccs.edu, blee@uccs.edu, j.ryoo@fdu.edu}
\thanks{Yongchan Jung and Shafayat Mowla Anik are graduate students.}
}

\newcommand{\SA}[1]{\textcolor{red}{[SA] #1}}

\maketitle
\begin{abstract}
Large language model inference on processing-in-memory (PIM) architectures promises to break the memory wall by performing multiply-accumulate (MAC) operations directly within HBM3 DRAM banks. Prior work identifies the power constraint timing parameter \nCCDAB{} as the primary performance bottleneck and optimizes scheduling accordingly. 
We demonstrate that for GEMV operations that dominate autoregressive decoding, the DRAM row cycle time (\nRC) is 10 to 11 times larger than \nCCDAB. Consequently, \nCCDAB{} is entirely masked, rendering prior \nCCDAB-focused optimizations ineffective for these workloads.
The root cause is inherited host-centric address interleaving, which forces every all-bank MAC command into a different DRAM row. We propose RH+ scheduling, a simple stride change that keeps 32 consecutive MAC operations within the same row. Cycle-accurate simulation across four LLM workloads shows that RH+ delivers 8--12$\times$ speedup, over 74\% energy reduction, and up to 52$\times$ EDP improvement.
\end{abstract}


\section{Introduction}
\label{sec:introduction}

\begin{figure}[t]
    \centering
    \includegraphics[width=\columnwidth]{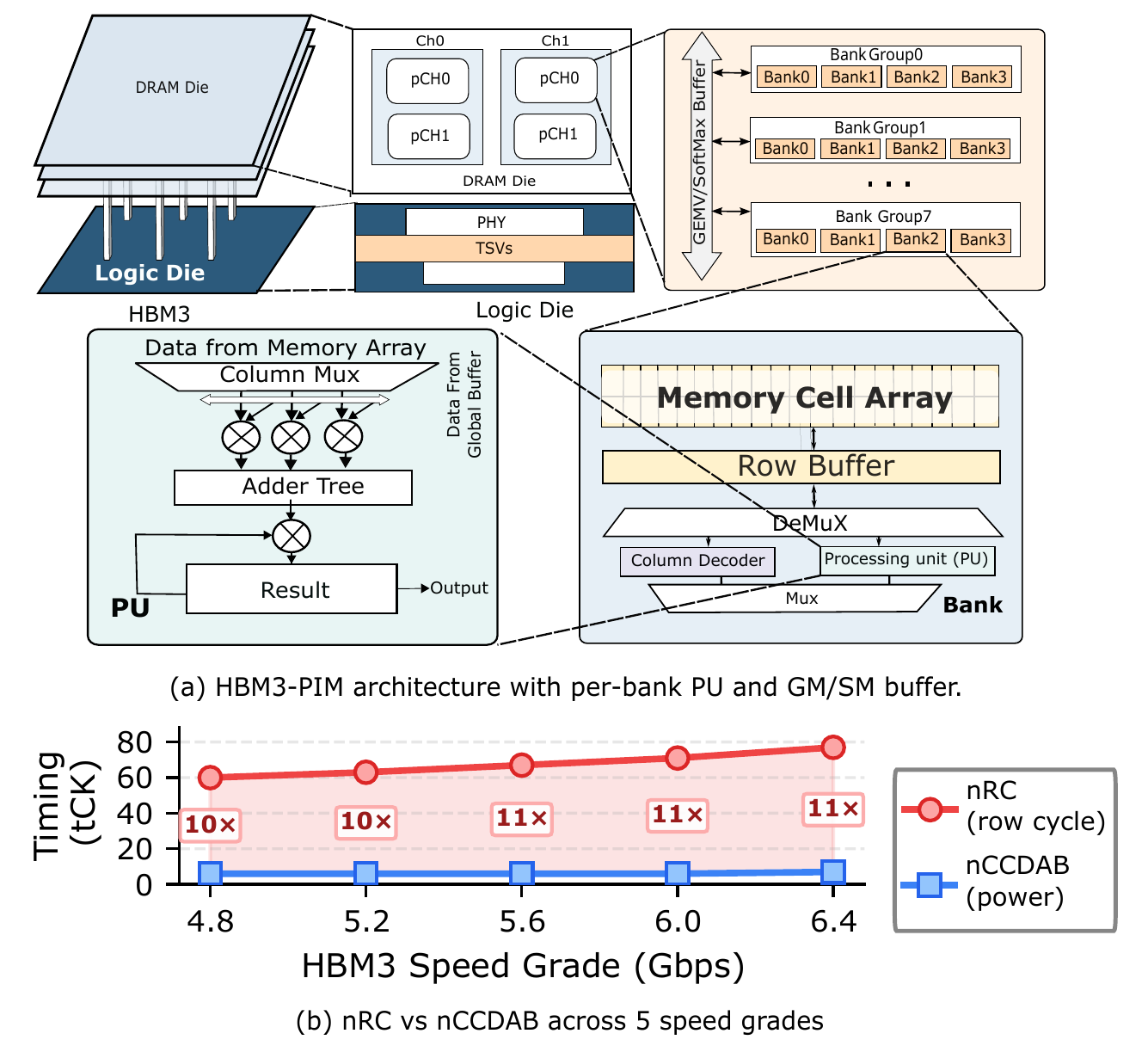}
    \caption{HBM3-PIM bank architecture and \nRC{} vs.\ \nCCDAB{} across five HBM3 speed grades~\cite{jedec_hbm3}.}
    \label{fig:motivation}
\end{figure}

Large language models (LLMs) have scaled from hundreds of millions to hundreds of billions of parameters in just a few years~\cite{gpt3, llama, megatron, opt}. While training these models remains compute-bound, serving them at low batch sizes tells a different story. During autoregressive decoding, each generated token requires reading the full weight matrix for every linear layer, yet performs only a single matrix-vector multiplication (GEMV) per layer. This memory-bound nature has motivated processing-in-memory (PIM) architectures that place compute units directly within the DRAM die, eliminating the costly data movement that conventional accelerators must endure~\cite{hbm2pim, hbmpim}. HBM3-PIM extends this idea to high-bandwidth memory stacks, enabling all-bank multiply-accumulate (MAC\_AB) operations that activate compute across all 1024 banks simultaneously~\cite{jedec_hbm3} as shown in Fig.~\ref{fig:motivation}(a).

Recent PIM-based LLM inference work~\cite{attacc, neupims} focuses on the power constraint timing parameter \nCCDAB, treating it as the primary performance bottleneck and scheduling around it. We observe, however, that a larger and previously overlooked bottleneck exists. As Fig.~\ref{fig:motivation}(b) shows, the DRAM row cycle time \nRC{} is 10 to 11 times larger than \nCCDAB{} across all HBM3 speed grades, completely masking the power constraint. The root cause, detailed in \S\ref{sec:background}, is a host-centric address interleaving stride that forces every MAC\_AB into a different DRAM row.

We propose \emph{RH+ scheduling}, which changes the MAC\_AB address stride from 64 columns to 1 column, enabling 32 consecutive row hits per activate-precharge cycle. This paper makes three contributions:
\begin{itemize}
    \item We identify \nRC, not \nCCDAB, as the true GEMV bottleneck in HBM3-PIM inference, showing that the power constraint is entirely masked by row management overhead.
    \item We propose RH+ scheduling, a stride change that converts every MAC from a row miss into a row hit, yielding an analytical speedup of 8.1--10.8$\times$.
    \item We validate RH+ end-to-end with cycle-accurate Ramulator~2.0 simulation~\cite{ramulator2} across four LLM models, demonstrating 8.25--11.88$\times$ speedup, 74.5--77.1\% energy reduction, and 32.4--52.0$\times$ EDP improvement.
\end{itemize}

\section{Background}
\label{sec:background}

\noindent\textit{HBM3-PIM Architecture:}
An HBM3-PIM stack consists of 16 channels, each containing 2 pseudo-channels and 2 ranks. Each rank is organized into 4 bank groups of 4 banks, yielding 1024 banks per stack~\cite{jedec_hbm3}. As shown in Fig.~\ref{fig:motivation}(a), every bank contains a processing unit (PU) capable of executing MAC operations on data resident in that bank's DRAM arrays. The all-bank MAC command (MAC\_AB) activates PUs across all banks simultaneously for data-parallel GEMV execution. Additional PIM commands handle inter-bank data movement and aggregation, including WR\_GB (write to global buffer, 4\,tCK), MV\_SB (move via shared buffer, 4\,tCK), and reduction operations (MV\_GB, SFM).

\noindent\textit{DRAM Row Management:}
As illustrated in Fig.~\ref{fig:motivation}(a), DRAM arrays are organized into rows and columns. Before any computation, the target row must be \emph{activated} (ACT), which copies the row contents into the sense amplifier (row buffer). After the operation completes, the row must be \emph{precharged} (PRE) to restore the bitlines. The minimum time from one ACT to the next on the same bank is the row cycle time, \nRC. Across HBM3 speed grades ranging from 4.8 to 6.4\,Gbps, \nRC{} spans 60 to 77\,tCK~\cite{jedec_hbm3}.

\noindent\textit{Power Constraint (\nCCDAB):}
To limit simultaneous switching current across all banks, the HBM3-PIM specification defines \nCCDAB{} as the minimum interval between consecutive MAC\_AB commands. Under the power-constrained (PC) mode, \nCCDAB{} ranges from 6 to 7\,tCK across speed grades. Systems with sufficient power delivery and thermal headroom may operate in non-power-constrained (NPC) mode, which relaxes \nCCDAB{} to 4\,tCK for higher throughput.

\begin{figure}[t]
    \centering
    \includegraphics[width=\columnwidth]{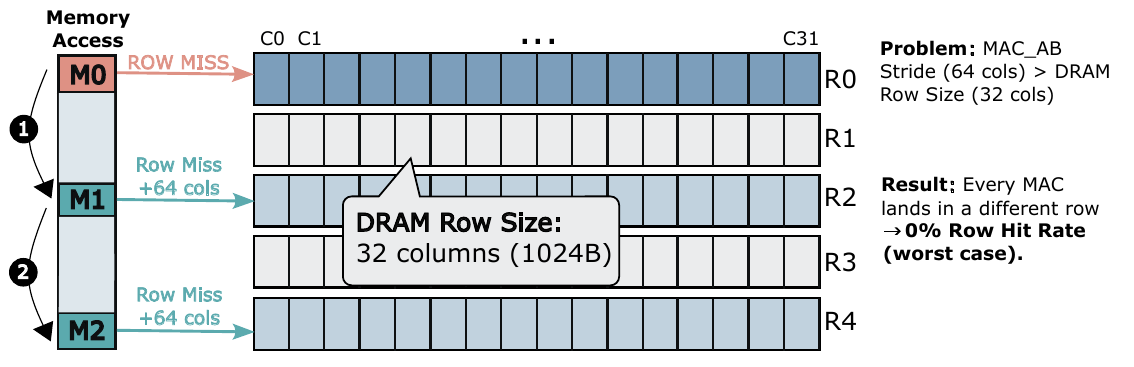}
    \caption{Baseline stride-64 row addressing. Each MAC\_AB lands in a different row, resulting in 0\% row hit rate. }
    \label{fig:row_addressing}
\end{figure}

\noindent\textit{DRAM Row Addressing and Bank-Rotation Stride:}
Standard HBM address interleaving maps consecutive cache-line addresses across different banks to maximize memory bus utilization for host-side access. This interleaving produces a per-bank address stride of 64 columns (2048\,B), meaning that consecutive addresses mapping to the \emph{same} bank are 64 columns apart. Prior PIM work~\cite{attacc} inherits this mapping for MAC\_AB commands without modification.

Fig.~\ref{fig:row_addressing} illustrates the consequence. Each DRAM row holds 32 columns (C0--C31, 1024\,B). The first MAC command M0 accesses a column in row R0. Because the stride is 64 columns, the next command M1 lands 64 columns later, which skips entirely past R1 and into R2. The third command M2 advances another 64 columns into R4. Every MAC therefore lands in a different row, forcing a full ACT-PRE cycle each time and yielding a 0\% row hit rate. Since MAC\_AB operates entirely within each bank and data never traverses the memory bus, this stride provides no benefit for PIM yet forces every MAC into a different row.

\begin{figure}[t]
    \centering
    \includegraphics[width=\columnwidth]{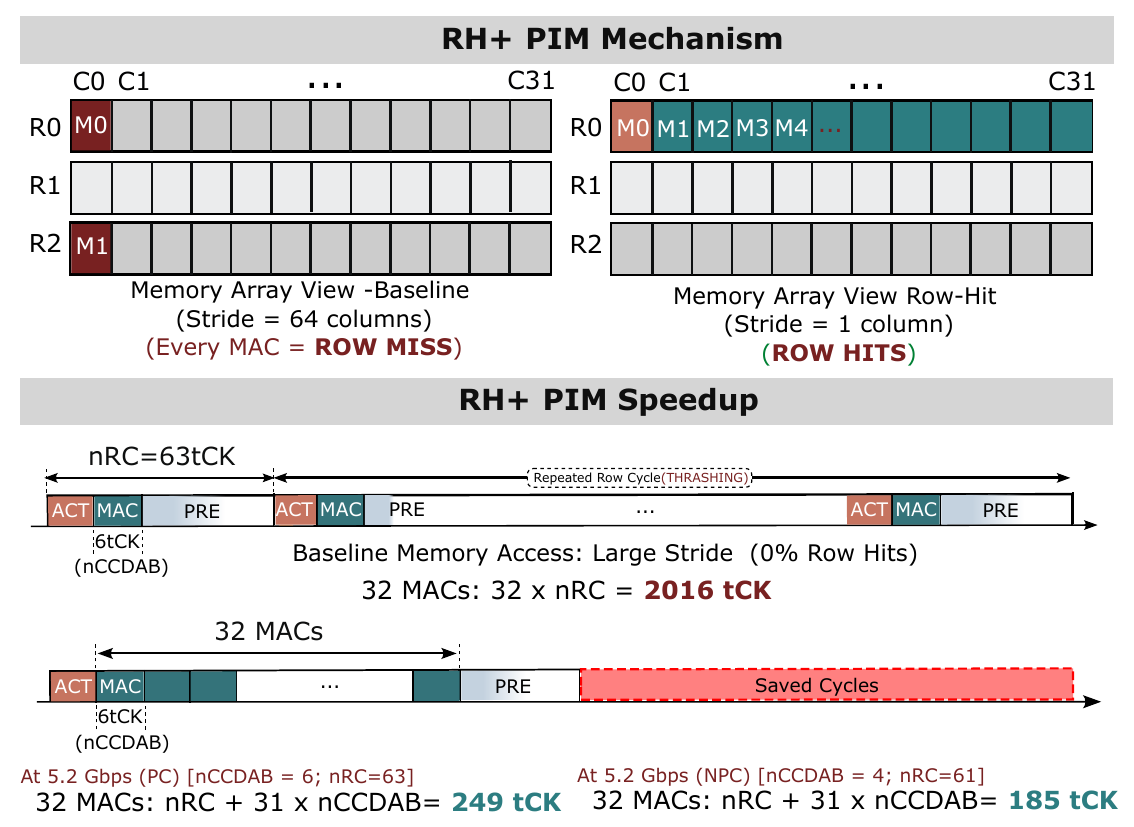}
    \caption{RH+ mechanism (top) and timing-level speedup (bottom) at the 5.2\,Gbps speed grade. }
    \label{fig:rowhit_mechanism}
\end{figure}

\section{The \nRC{} Bottleneck \& RH+ Optimization}
\label{sec:nrc_bottleneck}

\subsection{GEMV Dominance in Decode}
\label{subsec:gemv_dominance}

In autoregressive LLM decoding, each token generation requires a full forward pass through every transformer layer. At batch size~1, each layer performs GEMV operations for the QKV projection, output projection, and two feed-forward layers. These linear layers account for the vast majority of decode compute, with prior work reporting that GEMV operations consume over 95\% of per-token latency~\cite{attacc, neupims}. Our cycle-accurate simulation also confirms that compute operations consume 90+\% of total cycles across all model configurations, while rearrangement (data movement between buffers) contributes a negligible fraction. The fundamental inefficiency lies in how each MAC\_AB command interacts with the DRAM row buffer. In the baseline configuration, every MAC\_AB triggers a full row cycle: activate the target row (ACT), perform the computation over \nCCDAB\,=\,6\,tCK, then precharge the bank (PRE). The entire sequence takes \nRC\,=\,63\,tCK at 5.2\,Gbps, yielding a MAC duty cycle of only $6/63 = 9.5\%$.

\subsection{Power Constraint Is Irrelevant for GEMV}
\label{subsec:power_irrelevant}

Since every MAC\_AB incurs the full \nRC{} penalty (Section~\ref{sec:background}), \nCCDAB{} is entirely masked by row management. Even if \nCCDAB{} were reduced to zero, GEMV latency would remain unchanged. In HBM3-PIM, the 16-channel interleaving architecture means that MAC\_AB commands cycle through 16 channels before returning to the same channel, creating a natural inter-command gap of 16\,tCK per channel. This gap exceeds \nCCDAB{} at all speed grades (6--7\,tCK in PC, 4\,tCK in NPC). As a result, PC and NPC modes produce \emph{identical} cycle counts and energy consumption for baseline GEMV. Our kernel measurements at 5.2\,Gbps confirm this: both modes yield exactly 221K cycles for a GPT-175B QKV operation ($M{\times}K = 4608{\times}12288$ projection weight matrix).

\subsection{RH+ Scheduling}
\label{subsec:rowhit_scheduling}

Our key observation is that if consecutive MAC\_AB commands stayed within the same DRAM row, the costly ACT-PRE cycle would be amortized across multiple operations. RH+ scheduling achieves this by changing the MAC\_AB address stride from 64 columns to 1 column. Fig.~\ref{fig:rowhit_mechanism} illustrates the full mechanism and its performance impact. In the memory array view (Fig.~\ref{fig:rowhit_mechanism}, top), the left side recaps the baseline stride-64 row-miss pattern from Fig.~\ref{fig:row_addressing}. The right side shows RH+ with stride-1: M0, M1, M2, M3, M4, and subsequent MACs all access consecutive columns C0, C1, C2, \ldots\ within the same row R0. Because a single DRAM row holds 32 columns, RH+ keeps all 32 MAC\_AB commands within one row before the next ACT-PRE cycle is needed.
In the timing comparison (Fig.~\ref{fig:rowhit_mechanism}, bottom), the baseline repeats the full ACT-MAC-PRE cycle for each MAC, costing $32 \times \nRC = 2016$\,tCK. Under RH+, only a single ACT is issued at the start, the 32 MACs fire back-to-back at the \nCCDAB{} rate, and a single PRE closes the row at the end. The saved cycles are highlighted in red. The total drops to $\nRC + 31 \times \nCCDAB = 249$\,tCK in PC and 185\,tCK in NPC.

\begin{figure}[t]
    \centering
    \includegraphics[width=\columnwidth]{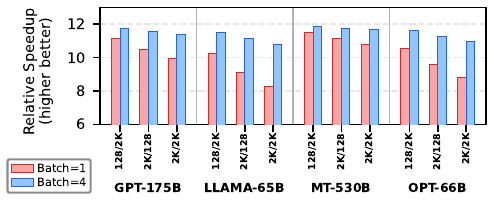}
    \caption{End-to-end speedup of RH+ over baseline.}
    \vspace{-2mm}
    \label{fig:speedup}
\end{figure}

\section{Methodology}
\label{sec:methodology}

We extend the AttAcc simulator~\cite{attacc}, which models HBM3-PIM command scheduling and end-to-end LLM inference, with Ramulator~2.0~\cite{ramulator2} for cycle-accurate DRAM timing with per-command-type cycle and energy tracking. The energy model follows the IDD-based DRAMPower methodology~\cite{drampower}, decomposing total energy into four components. \emph{Compute energy} accounts for MAC\_AB operations. \emph{Rearrangement energy} covers data movement commands such as WR\_GB, MV\_SB, MV\_GB, and SFM. \emph{ACT/PRE/REF energy} captures the cost of row management and refresh. Finally, \emph{background energy} reflects standby currents IDD3N (active) and IDD2N (idle) across all 32 pseudo-channels for the full execution duration.

We evaluate four representative LLM models under tensor-parallel inference (tp\,=\,8) across five HBM3 stacks: GPT-175B~\cite{gpt3} (96 layers, 96 heads, $d_\text{model}$\,=\,12288), LLaMA-65B~\cite{llama} (80 layers, 64 heads, $d_\text{model}$\,=\,8192), Megatron-Turing 530B~\cite{megatron} (105 layers, 128 heads, $d_\text{model}$\,=\,20480), and OPT-66B~\cite{opt} (64 layers, 72 heads, $d_\text{model}$\,=\,9216). Each model is evaluated with three input/output sequence length pairs (128/2048, 2048/128, 2048/2048) and two batch sizes (1 and 4), yielding 24 configurations (6 per model), each evaluated under both baseline and RH+ scheduling. All experiments use the HBM3 5.2\,Gbps speed grade.


\section{Evaluation}
\label{sec:evaluation}

\begin{figure}[t]
    \centering
    \includegraphics[width=\columnwidth]{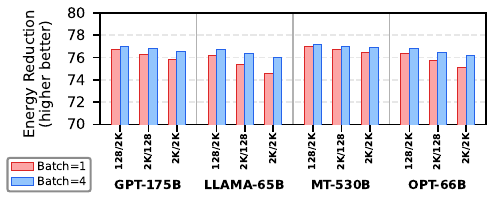}
    \caption{End-to-end energy reduction (\%) of RH+.}
    \label{fig:energy}
\end{figure}

\subsection{End-to-End Performance}
\label{subsec:e2e_speedup}

Fig.~\ref{fig:speedup} shows the speedup achieved by RH+ scheduling across all configurations. At batch\,=\,1, speedup ranges from 8.25$\times$ (LLaMA-65B, 2048/2048) to 11.51$\times$ (MT-530B, 128/2048). At batch\,=\,4, the range narrows and rises to 10.77--11.88$\times$. Two properties of LLM workloads explain this variation. First, the speedup correlates with the model's hidden dimension $K$, which determines the number of MAC\_AB commands per GEMV layer. MT-530B ($K$\,=\,20480) issues the most MACs per layer, so the row-hit benefit of RH+ accumulates over more operations, yielding the highest speedup. In contrast, LLaMA-65B and OPT-66B ($K$\,=\,8192 and 9216) issue fewer MACs, and the fixed overhead of attention, which RH+ does not optimize, occupies a relatively larger fraction of total time. Second, batch size has a pronounced effect. At batch\,=\,1, each token's forward pass includes both GEMV (weight projection) and attention (KV cache lookup). Attention latency is independent of RH+ because it does not use MAC\_AB commands. Longer output sequences (Lout\,=\,2048) further increase attention's share, pulling speedup down. At batch\,=\,4, however, each batch element contributes an additional full GEMV while the attention cost remains per-sequence. This shifts the compute mix heavily toward GEMV, narrowing the speedup range across models.

\begin{figure}[h]
    \centering
    \includegraphics[width=\columnwidth]{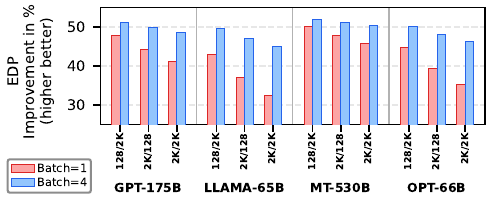}
    \caption{End-to-end EDP improvement of RH+.}
    \label{fig:edp}
\end{figure}
\vspace{-2mm}

\subsection{End-to-End Energy}
\label{subsec:e2e_energy}

Fig.~\ref{fig:energy} shows energy reduction across the same configurations. Savings range from 74.5\% to 77.1\%, a remarkably tight band considering the 3.3$\times$ variation in model size. The reason lies in the HBM3-PIM power structure. An HBM3 stack contains 32 pseudo-channels, each drawing standby current (IDD3N in active mode, IDD2N in idle mode) for the entire execution duration regardless of how many banks are computing. This background power is fixed per unit time and cannot be clock-gated during GEMV. In the baseline, the long \nRC-dominated execution stretches this standby cost over many cycles, making background energy the dominant component (roughly 67\% of total). When RH+ reduces execution time by roughly 10$\times$, the standby duration shrinks proportionally, and background energy collapses. Because standby current is a DRAM property rather than a workload parameter, the energy reduction is nearly model-independent.

\subsection{End-to-End Energy-Delay Product}
\label{subsec:e2e_edp}

Fig.~\ref{fig:edp} shows Energy-Delay Product (EDP) improvement ranging from 32.4$\times$ to 52.0$\times$. EDP\,=\,Time\,$\times$\,Energy captures the compounding relationship between performance and energy, and the superlinear gains arise directly from the background energy structure described above. When RH+ reduces execution time, the time reduction enters EDP twice: once through the delay term and again through the energy term, because background energy is itself proportional to execution time. Concretely, a 10$\times$ speedup with 76\% energy reduction yields roughly $10/(1{-}0.76) \approx 41.7\times$ EDP improvement. MT-530B at batch\,=\,4 reaches the peak of 52.0$\times$ because its large $K$ maximizes speedup (11.88$\times$) and the high GEMV share maximizes energy reduction (77.1\%) simultaneously. 

\begin{figure}[t]
    \centering
    \includegraphics[width=\columnwidth]{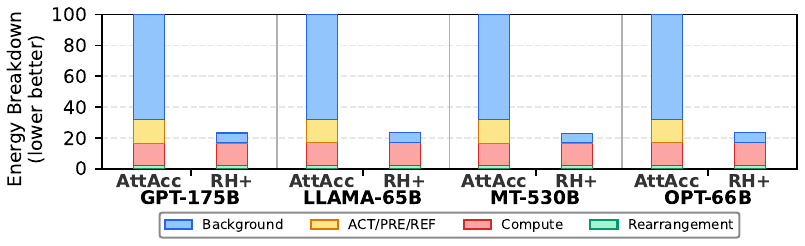}
    \caption{Energy breakdown normalized to baseline.}
    \label{fig:energy_breakdown}
\end{figure}

\subsection{Energy Breakdown}
\label{subsec:energy_breakdown}

Fig.~\ref{fig:energy_breakdown} decomposes energy into four components, normalized to the baseline total for each model. In the baseline, background energy dominates at 67.3--67.8\%, followed by ACT/PRE/REF overhead (15.5\%), compute (14\%), and rearrangement (3\%). The large ACT/PRE/REF share is a direct consequence of the 0\% row-hit rate in baseline: every MAC\_AB pays the full activate-precharge cost. After RH+, this component nearly vanishes because 31 out of every 32 MACs are row hits that skip ACT/PRE entirely. Background energy also collapses in absolute terms due to the shorter execution, but its \emph{relative} share actually rises to 25--31\% of the now much smaller total, because compute energy (which is the same number of MAC operations regardless of scheduling) becomes the dominant remaining cost. 

\section{Discussion \& Conclusion}
\label{sec:conclusion}

\noindent\textbf{Discussion:}
RH+ requires a one-time offline reordering of weight elements along the K-dimension so that consecutive elements map to adjacent DRAM columns within the same row. This transformation has zero runtime overhead and does not change the GEMV result, as accumulation order is commutative.
After RH+ eliminates the \nRC{} bottleneck, \nCCDAB{} becomes the new limiting factor. PC (\nCCDAB\,=\,6) is 1.36$\times$ slower than NPC (\nCCDAB\,=\,4) under RH+, compared to identical performance in the baseline. Combining RH+ with NPC-aware scheduling is a promising direction for future work.

\noindent\textbf{Related Work:}
HBM-PIM~\cite{hbm2pim}, AIM~\cite{aim}, Newton~\cite{newton}, and TransPIM~\cite{transpim} target PIM-based neural network inference. AttAcc~\cite{attacc} and NeuPIMs~\cite{neupims} specifically address LLM inference, optimizing around the power constraint. None identify \nRC{} as the GEMV bottleneck. RH+ is complementary to these approaches.

\noindent\textbf{Conclusion.}
We have shown that \nRC, not \nCCDAB, is the dominant bottleneck for PIM-based LLM inference. RH+ achieves 8.25--11.88$\times$ speedup, 74.5--77.1\% energy reduction, and 32.4--52.0$\times$ EDP improvement across four LLM models, demonstrating that DRAM microarchitectural awareness is essential for realizing the full potential of PIM.

\bibliographystyle{IEEEtran}
\bibliography{refs}

\end{document}